\documentclass[runningheads]{llncs}
\usepackage{graphicx}
\usepackage{mathtools}
\usepackage{listings}
\usepackage{proof}
\usepackage{amsfonts}
\usepackage{amssymb}
\usepackage{stmaryrd}
\usepackage{algorithm,algorithmic}

\begin{document}
\title{Sunspot: A Decentralized Framework Enabling Privacy for Authorizable Data Sharing on Transparent Public Blockchains\thanks{This is the preprint version of the conference paper "Sunspot: A Decentralized Framework Enabling Privacy for Authorizable Data Sharing on Transparent Public Blockchains" in \textit{Proc. International Conference on Algorithms and Architectures for Parallel Processing, 2021}.}}
\titlerunning{Sunspot}
%
\author{Yepeng Ding \and
Hiroyuki Sato}
\authorrunning{Y. Ding and H. Sato}
%
\institute{The University of Tokyo, Tokyo, Japan\\
\email{\{youhoutei,schuko\}@satolab.itc.u-tokyo.ac.jp}}
\maketitle              
\begin{abstract}
Blockchain technologies have been boosting the development of data-driven decentralized services in a wide range of fields. However, with the spirit of full transparency, many public blockchains expose all types of data to the public such as Ethereum. Besides, the on-chain persistence of large data is significantly expensive technically and economically. These issues lead to the difficulty of sharing fairly large private data while preserving attractive properties of public blockchains. Although direct encryption for on-chain data persistence can introduce confidentiality, new challenges such as key sharing, access control, and legal rights proving are still open. Meanwhile, cross-chain collaboration still requires secure and effective protocols, though decentralized storage systems such as IPFS bring the possibility for fairly large data persistence. In this paper, we propose Sunspot, a decentralized framework for privacy-preserving data sharing with access control on transparent public blockchains, to solve these issues. We also show the practicality and applicability of Sunspot by MyPub, a decentralized privacy-preserving publishing platform based on Sunspot. Furthermore, we evaluate the security, privacy, and performance of Sunspot through theoretical analysis and experiments.

\end{abstract}
\section{Introduction}
\label{sec:intro}
Worldwide decentralization of data persistence and sharing is advancing with the evolution of public blockchain technologies, as can be seen by the boom of decentralized applications (DApp), especially decentralized finance (DeFi) \cite{zetzsche_decentralized_2020} in recent years. Different from decentralized payment platforms such as Bitcoin \cite{nakamoto_bitcoin_2019} \footnote{For brevity, when mentioning the name of a cryptocurrency in this paper, we refer to its underlying blockchain. For instance, “Bitcoin” refers to the blockchain built on the Bitcoin Backbone Protocol \cite{garay_bitcoin_2015}.}, DeFi has wider applicability that provides various financial products and services such as lending \cite{bartoletti_sok_2021}, insurance \cite{wan_pride_2018}, trading, and investment \cite{chen_blockchain_2020}. Although the energy consumption of DeFi compared to traditional finance is still under debate, DeFi has presented a set of attractive properties including accessibility, automation, transparency, interoperability, finality, borderlessness, and innovativeness. These properties are generally achieved through on-chain data persistence and sharing supported by transparent public blockchains such as Ethereum \cite{wood_ethereum_2014} that are permissionless and fully disclosing all types of data. But it would still be limited if data are finance-oriented only.

With the maturity of the ERC-721 \cite{entriken_erc-721_2018}, the application based on non-fungible token (NFT) techniques has enlarged applicability to all types of digital assets in a wide range of fields including art, law, finance, entertainment, as well as personal data. An NFT is generally used to represent legal rights such as ownership of a digital asset, though is possible for a traditional asset. The proofs of these legal rights possess beneficial properties of public blockchains such as tamperproofing, transparency, traceability, and high availability due to its nature of solidifying proof data on public blockchains. However, proof of legal rights does not imply prevention of infringement of these rights because off-chain assets may still face security and privacy issues.

Although on-chain storage of digital assets is theoretically feasible, it is too expensive to store large data directly on public blockchains due to the consensus difficulty. Consequently, current public blockchains have strict data size limitations. While widespread centralized solutions such as self-hosted storage and cloud storage \cite{wu_cloud_2010} have lower costs, some properties like integrity and availability are hard to implement. As a compromise solution, public blockchains that leverage peer-to-peer file sharing protocols have been experimented in industry to store large data without centralized mechanisms and authorities such as InterPlanetary File System (IPFS) \cite{benet_ipfs-content_2014}. With elaborated incentive mechanisms and cryptographic support, some blockchains have made significant progress to attract participants such as the Filecoin network that is built on top of IPFS. These systems enhance the security of data storage including integrity, durability, and availability but still do not solve privacy issues because all data are stored publicly without any confidentiality protection at default. In other words, anyone with a public link can fully acquire that data. Even if data are encrypted by data providers before storage, it is challenging to share decryption keys and authorize data consumers without central entities.

From the background above, we summarize and formalize challenges of decentralized data sharing as follows, where $\textbf{D}$ is the shared data set, $\textbf{R}$ the legal right set, $\textbf{U}$ the user set.

\paragraph{Challenge 1}[Proof of legal rights]
$\mathcal{P}_r(d, r) = u$ if and only if $u$ has the legal right $r$, where $\mathcal{P}_r: \textbf{D} \times \textbf{R} \to \textbf{U}, d \in \textbf{D}, r \in \textbf{R}$.

\paragraph{Challenge 2}[Fairly large data storage]
$\exists d \in \textbf{D}, \textit{Size}(d) > 1 \text{MB}$.

\paragraph{Challenge 3}[Interoperability]
The system is compatible with various blockchains and allows multiple heterogeneous blockchains to implement functionalities with minimum compatibility issues.

\paragraph{Challenge 4}[Privacy preservation]
$\forall d \in \textbf{D}$, $d$ is encrypted.

\paragraph{Challenge 5}[Access control]
$\mathcal{P}_a(u,d) = \textit{Dec}(d)$ if and only if $u$ has the access privilege of $d$, where $\mathcal{P}_a: \textbf{U} \times \textbf{D} \to \{ \textit{Dec}(d): d \in \textbf{D} \}$.

Motivated by the above issues and challenges, we propose a framework named Sunspot to enable privacy and provide access control for flexible data sharing on transparent public blockchains. Sunspot has a decentralized architecture and supports multiple access control mechanisms based on customizable identity management models. We generalize its protocols to make Sunspot compatible with various public blockchains and decentralized storage systems. Besides, Sunspot has a cross-chain collaboration mechanism that bridges blockchains and decentralized storage systems, as well as a multi-chain mechanism that allows heterogeneous blockchains to present unified characteristics.

We summarize our main contributions of this paper and Sunspot as follows.
\begin{enumerate}
    \item We present Sunspot, a decentralized privacy-preserving data sharing framework that solves the challenges above. To the best of our knowledge, Sunspot is the first framework that focuses on preserving privacy for data sharing on transparent public blockchains.
    \item We generalize protocols and provide support for multiple access control mechanisms to enable flexible data sharing in various scenarios. Particularly, we show an identity management model and a management-free model for identification based on two different access control mechanisms.
    \item We present MyPub, a decentralized privacy-preserving publishing platform based on Sunspot, to show the practicality and applicability of Sunspot.
    \item We prove the security and privacy properties of Sunspot and demonstrate them with experiments from the perspective of attackers.
    \item We evaluate the performance of core functionalities of Sunspot.
\end{enumerate}

\section{Related Work}
The privacy issues coming with full transparency are rising concerns in society and endangering legacy of some decentralized services due to tightening of regulations such as the General Data Protection Regulation (GDPR) \cite{regulation_regulation_2016}. In such a context, many enterprise products turn into solutions based on private blockchains and consortium blockchains such as HyperLedger Fabric \cite{androulaki_hyperledger_2018} and XRP Ledger \cite{schwartz_ripple_2014}. In some scenarios, redeveloping an open-source public blockchain as a privacy-preserving blockchain also becomes a solution.

For existing transparent public blockchains, solutions based on zero-knowledge proof (ZKP) \cite{goldreich_definitions_1994} are attracting attention. As an early exploration, Zerocoin \cite{miers_zerocoin_2013} provides a feasible solution to Bitcoin’s pseudonymity by zero-knowledge proofs for set membership. Security such as prevention of double spending is assured without revealing the transaction. But the original implementation of Zerocoin significantly increased computation cost and slowed down the network efficiency. Later, Zcash, the first widespread blockchain based on the Zero-Knowledge Succinct Non-Interactive Argument of Knowledge (zk-SNARKs) \cite{groth_updatable_2018}, was launched on top of the Zerocoin protocol. Zcash enhances privacy by making the sender, recipient and amount completely private via encryption as well as improves network performance. However, it is merely a privacy protocol in the Bitcoin network that can neither support smart contracts nor store large data.

Besides ZKP-based approaches, Monero \cite{sun_ringct_2017}, which is based on the CryptoNote \cite{van_saberhagen_cryptonote_2013}, provides another solution based on Ring Confidential Transactions (RingCT) \cite{noether_ring_2015}. It uses a variant of linkable ring signature \cite{liu_linkable_2004} to allow a member of a group to stay in anonymity while signing messages on behalf of the group. However, the same with ZKP-based approaches, the smart contract is not supported. These privacy-preserving blockchains are generally used for decentralized payment only and not for extended functionalities such as generalized data sharing.

As for blockchain-based data sharing, many solutions have been proposed in different scenarios. For AI-powered network operations, Zhang et al proposed a mutual trust data sharing framework \cite{zhang_blockchain-based_2018} to break data barriers between different operators. This framework contains three layers: system management layer, storage layer, and user layer. Only the system management layer adopts a blockchain for control, while large data are stored in the cloud. Consequently, it is still vulnerable to threats to cloud storage and threatened by the trustiness of cloud service providers. In work \cite{xia_bbds_2017}, a framework named BBDS is proposed for data sharing of electronic medical records (EMR). It adopts a decentralized access control mechanism to secure medical records. However, BBDS is based on a permissioned blockchain and also relies on cloud storage. And privacy issues are not addressed including data encryption and key sharing. Recently, Derepo \cite{ding_derepo_2020}, a distributed data repository, is proposed to preserve the privacy of medical data with decentralized access control and homomorphic encryption (HE). Although the HE scheme proves to be effective for solving privacy issues, it leads to a significant performance issue.

To the best of our knowledge, there is no existing work that tackles the privacy issues for decentralized and general-purpose data sharing on transparent public blockchains.

\section{Sunspot}
\label{sec:sunspot}
Sunspots are dark areas on the surface of the Sun. Analogically, our framework enables private data sharing "in the dark", i.e., without disclosing any meaningful information from the shared data to the unauthorized public, on fully transparent public blockchains. Meanwhile, it is built on top of public blockchains to preserve their security properties.

Note, in the rest of the paper, we simply use the word \textit{blockchain} when referring to the phrase \textit{transparent public blockchain}. In Sunspot, we introduce new terminologies to clarify non-standardized concepts. We categorize blockchains into two types: control chain and storage chain. A blockchain can be a control chain if and only if it supports the smart contract, event mechanism, and NFT protocol. A storage chain is a blockchain that supports decentralized large data storage, which usually adopts a hybrid architecture. A blockchain can be both a control chain and a storage chain with the only conceptual difference. Notably, with the feature of the multi-chain compatibility of Sunspot, a control chain or a storage chain can be composed of a set of heterogeneous blockchains.

\subsection{Architecture}
Sunspot involves two main types of roles: organizer and participant. Organizers are responsible for initialization, deployment, and optionally for identification that is a continuous work to endorse participants’ identities. In this paper, we will show an identity management model based on a fine-grained access control mechanism in this section, a management-free model that depends on a payment-based access control mechanism in Section~\ref{sec:mypub}, and a self-sovereign identity model in Section~\ref{sec:discussion}.

Participants in Sunspot can be data providers, data consumers, or both. We call participants who are willing to share data as data providers while those who are willing to acquire data as data consumers. There is no direct interaction between a provider and a consumer. Sunspot encapsulates all operations and exposes interfaces to participants in a decentralized manner, which means it has a high probability that the interaction target is not the intended target but a set of unknown nodes in the blockchain network that are even not aware of the functionality.

\begin{figure}
\includegraphics[width=\textwidth]{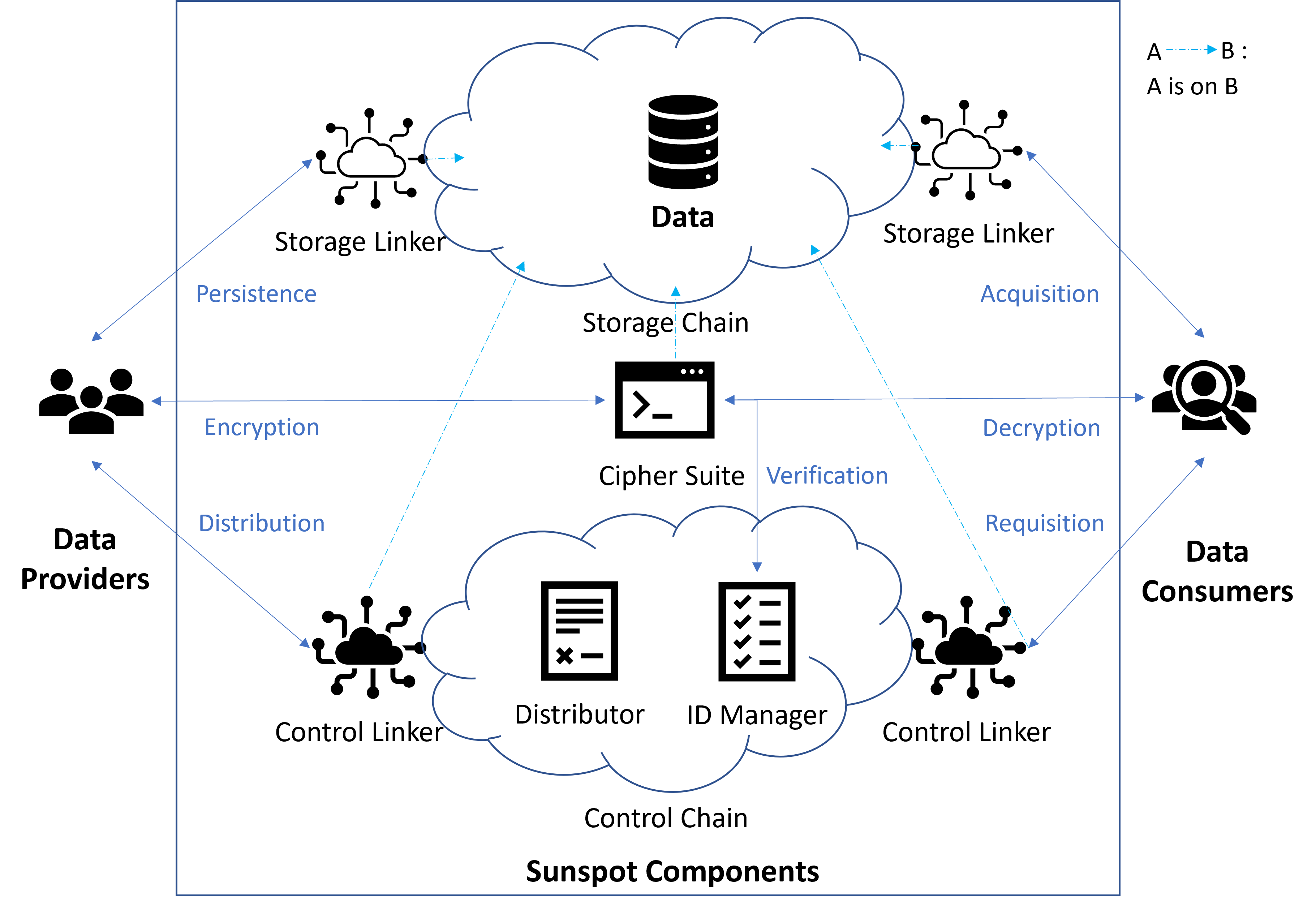}
\caption{The architecture of Sunspot.}
\label{fig:architecture}
\end{figure}

As shown in Figure~\ref{fig:architecture}, Sunspot contains five components: Distributor, ID Manager, Cipher Suite, Control Linker, and Storage Linker. From the view of deployment, except Distributor and ID Manager that are deployed on the control chain, other components are initialized on the storage chain.

\subsubsection{Distributor}
Distributor is a smart contract on the control chain that implements the NFT protocol. It provides core functionalities to solidify the legal rights of data by minting a unique NFT as its meta-information. A data owner uses Distributor to share data while a data consumer to query and get the meta-information of the shared data.

\subsubsection{ID Manager}
ID Manager is used for participant registration and identity query, which is a smart contract on the control chain. Participant registration is the key to integrating identification information such as wallet addresses of different blockchains and authorization information. For the identity management model in this section, Ciphertext-Policy Attribute-Based Encryption (CP-ABE) \cite{waters_ciphertext-policy_2011} is used for fine-grained access control. Hence, authorization information is a private key that describes the attributes of a participant. And identity query is a read-only operation for verification.

\subsubsection{Cipher Suite}
Cipher Suite plays a pivotal role in enabling privacy and securing access control. Its main functionalities are encrypting data and generating decryptors. It is designed to be secure and implemented with a set of security measurements such as obfuscation \cite{goldwasser_best-possible_2007} to ensure control-flow integrity (CFI) \cite{abadi_control-flow_2009}, type safety, and memory safety.

\subsubsection{Control Linker}
Control Linker is the gateway to the control chain that encapsulates interactions such as signing transactions and making the remote procedure call (RPC). A common interface is implemented to facilitate the redevelopment and multi-chain support.

\subsubsection{Storage Linker}
Similar to the Control Linker, Storage Linker is the gateway to the storage chain. Particularly, it encapsulates interactions with the file system such as uploading data to and downloading data from the storage chain.

\subsection{Protocol}
\label{sec:protocol}

\subsubsection{Environment Assumption}
\label{sec:protocol_env}
We assume there is an organizer set $\textbf{O}$ and a participant set $\textbf{P}$. In the worst case, $g \leq |\textbf{O}|$ of organizers are available, and at least one organizer of the available ones is honest and never colludes with the others.

The control chain $\mathfrak{C}$ and the storage chain $\mathfrak{S}$ already exist. For clarity, we assume $\mathfrak{C}$ is merely a blockchain. And so is $\mathfrak{S}$. We will illustrate the multi-chain case in Section~\ref{sec:multi_chain}.

Blockchains selected for $\mathfrak{C}$ and $\mathfrak{S}$ enable Transport Layer Security (TLS) for the JSON-RPC request/response communication channel.

Two master key pairs used to initialize Sunspot, $(\text{MPK}^c, \text{MSK}^c, \text{MWA}^c)$ for interacting with $\mathfrak{C}$, $(\text{MPK}^s, \text{MSK}^s, \text{MWA}^s)$ for interacting with $\mathfrak{S}$, are created, where $\text{MPK}$ is a master public key, $\text{MSK}$ a master secret key, and $\text{WMA}$ a master wallet address. Besides, $\text{MSK}^c$ and $\text{MSK}^s$ are encoded as integers in $\mathbb{Z}/m\mathbb{Z}$ and divided into $|\textbf{O}|$ shares respectively based on the Shamir's Secret Sharing (SSS) \cite{shamir_how_1979} with $(g, |\textbf{O}|)$-threshold. And each share for each of them is distributed to a unique organizer.

All participants already have key pairs for interacting with $\mathfrak{C}$ and $\mathfrak{S}$. $\forall P_i \in \textbf{P}$, $P_i$ has a pair $(\text{PK}_i^c, \text{SK}_i^c, \text{W}_i^c)$ for $\mathfrak{C}$ and $(\text{PK}_i^s, \text{SK}_i^s, \text{WA}_i^s)$ for $\mathfrak{S}$, where $\text{PK}$, $\text{SK}$, $\text{WA}$, denote public key, secret key, and wallet address respectively. For Ethereum, $\text{SK}$ is generated as random 256 bits usually by SHA-256 \cite{dang_secure_2015}. $\text{PK}$ is derived from $\text{SK}$ by the Elliptic Curve Digital Signature Algorithm (ECDSA) \cite{johnson_elliptic_2001} for the elliptic curve secp256k1. And $\text{WA}$ is created by applying the Keccak-256 \cite{dworkin_sha-3_2015} to $\text{PK}$.

\subsubsection{Initialization}
This protocol is used for initializing Sunspot.
\begin{enumerate}
    \item Deploy Distributor and ID Manager contracts to $\mathfrak{C}$ with $\text{MSK}^c$ and get addresses $C_{dis}$ and $C_{id}$ as their identifiers.
    \item Run $\textit{Setup}(\lambda, \Gamma)$ algorithm of CP-ABE \cite{waters_ciphertext-policy_2011} to produce public parameter $\hat{\text{PK}}$ and a master key $\hat{\text{MK}}$, where $\lambda$ is the security parameter and $\Gamma$ is the attribute space;
    \item Store $\hat{\text{PK}}$ and $\hat{\text{MK}}$ to state variables of $C_{dis}$ by invoking $\textit{SetABE}(\hat{\text{PK}}, \hat{\text{MK}})$;
    \item Configure and compile Cipher Suite, Control Linker, and Storage Linker with parameters $C_{dis}$ and $C_{id}$. Store them to $\mathfrak{S}$ with $\text{MSK}^s$ and get addresses $S_{ciph}$, $S_{cl}$, $S_{sl}$ as their identifiers;
    \item Emit a $\textit{Storage Registered Event} (\text{MWA}^s, S_{ciph}, S_{cl}, S_{sl})$ on $\mathfrak{C}$ with $\text{MSK}^c$;
    \item Make $\hat{\text{PK}}$, $\hat{\text{MK}}$, $S_{ciph}$, $S_{cl}$, $S_{sl}$ public to participants.
\end{enumerate}

\subsubsection{Registration}
Users need to execute this protocol to get qualified as participants or rejected. Suppose an arbitrary user $u \in \textbf{U}$ requests to get registered as a participant.
\begin{enumerate}
    \item $u$ submits a structured record $\text{Rec}_u$ with proofs to organizers who are responsible for identity management. $\text{Rec}_u$ consists of $\text{WA}_u^c$, $\text{WA}_u^s$ and a set of attributes $\textbf{A}$;
    \item Run $\textit{KeyGeneration}(\hat{\text{MK}}, \textbf{A})$ algorithm of the CP-ABE to produce a private key $\chi_u$;
    \item A $\textit{Registered Event} (\text{WA}_u^c, \text{WA}_u^s, \chi_u)$ is emitted on $\mathfrak{C}$ by organizers via $C_{id}$ to indicate $u \in \textbf{P}$ if $\text{Rec}_u$ is approved.
\end{enumerate}

\subsubsection{Solidification}
Data providers use this protocol to solidify their data while preserving privacy through three sub-protocols: Encryption, Persistence, and Distribution.

Suppose a data provider $P_d \in \textbf{P}$ wants to share data $D$ in private.

\paragraph{Encryption.}
After obtaining Cipher Suite from $S_{ciph}$, execute this protocol to produce encrypted data $\textit{Enc}(D)$ and corresponding decryptor $\textit{Dec}$.
\begin{enumerate}
    \item Generate a secret key $\kappa = \textit{Bcrypt}(\text{SK}_d^c \oplus \epsilon)$ where $\textit{Bcrypt}$ is a password scheme \cite{provos_bcrypt_1999}, $\epsilon$ is a nonce that is generated from a cryptographically-secure pseudorandom number generator \cite{vazirani_efficient_1984}, and $\oplus$ denotes an operation to obfuscate $\text{SK}_d^c$ and $\epsilon$ such as concatenation and bitwise exclusive or;
    \item Get $\textit{Enc}(D)$ by encrypting $D$ with $\kappa$ based on the AES-GCM-SIV scheme (AES) \cite{gueron_aes-gcm-siv_2017};
    \item Run $\textit{Encrypt}(\hat{\text{PK}}, M, \gamma)$ algorithm of CP-ABE to produce an access control policy $\rho$, where $M$ is a random challenge message and $\gamma$ is an access structure over $\Gamma$;
    \item Generate decryptor $\textit{Dec}$ with $\kappa$, $\rho$ and $M$.
\end{enumerate}

\paragraph{Persistence.}
This sub-protocol is executed via $S_{sl}$ to store $\textit{Enc}(D)$ and $\textit{Dec}$ on $\mathfrak{S}$ and produces metadata address $S_\mu$ where storing the meta-information of $D$.
\begin{enumerate}
    \item Get persistent addresses $S_{data}$ and $S_{dec}$ after storing $\textit{Enc}(D)$ and $\textit{Dec}$ to $\mathfrak{S}$ with $\text{SK}_d^s$ signing transactions;
    \item Generate a structured metadata $\mu$ that at least contains $S_{data}$, $\mathcal{H}(\textit{Enc}(D))$, $S_{dec}$, and $\mathcal{H}(\textit{Dec})$ where $\mathcal{H}: \{ 0, 1 \}^* \to \{ 0, 1 \}^l$ is a cyrptographic hash function with a fixed size $l$.
    \item Get persistent address $S_\mu$ after storing $\mu$ to $\mathfrak{S}$ with $\text{SK}_d^s$.
\end{enumerate}

\paragraph{Distribution.}
This sub-protocol solidifies $S_\mu$ as an NFT on $\mathfrak{C}$ through $S_{cl}$. After this protocol, $P_d$ can make $S_\mu$ public to share $D$ in a decentralized and private way.
\begin{enumerate}
    \item Generate a unique token id $\iota$;
    \item Invoke $\textit{Mint}(\text{WA}_d^c, \iota)$, a method in the NFT protocol, of $C_{dis}$ with $\text{SK}_d^c$;
    \item Update ID mapping $\mathcal{M}_{id}$ with $S_\mu \mapsto \iota$;
    \item Emit a $\textit{Distributed Event} (\text{WA}_d^c, \iota)$ on $C_{dis}$.
\end{enumerate}

\subsubsection{Authorization}
Data consumers use this protocol to get authorized and acquire decrypted data through four sub-protocols: Requisition, Verification, Acquisition, and Decryption.

Suppose a data consumer $P_c \in \textbf{P}$ knows $S_\mu$ and wants to acquire decrypted data $D$.

\paragraph{Requisition.}
In this protocol, $P_c$ explicitly makes a request via $S_{cl}$. After this protocol, $P_c$ knows $\iota$.

\begin{enumerate}
    \item Validate identity existence by ensuring a $\textit{Registered Event} (\text{WA}_c^c, \text{WA}_c^s, *)$ exists, where $\text{WA}_c^c$ is derived from $\text{SK}_c^c$, $\text{WA}_c^s$ from $\text{SK}_c^s$. If the validation result is $\bot$, terminate the protocol with \textit{Non-existence} exception. Otherwise, continue the protocol;
    \item Get $\iota$ by $\mathcal{M}_{id}(S_\mu)$;
    \item Emit a $\textit{Requested Event} (\text{WA}_c^c, \iota)$ with $\text{SK}_c^c$.
\end{enumerate}

\paragraph{Acquisition.}
By this sub-protocol, $P_c$ gets decryptor $\textit{Dec}$ via $S_{sl}$.
\begin{enumerate}
    \item Get $\mu$ from $S_\mu$;
    \item Get $\textit{Dec}$ from $\mu[dec] \implies S_{dec}$ and verify its integrity by $\mathcal{H}(\textit{Dec})$.
\end{enumerate}

\paragraph{Verification.}
This sub-protocol is enforced by $\textit{Dec}$ in memory to authorize a request and produce verification result $\eta \in \{ \top, \bot \}$.

\begin{enumerate}
    \item Validate identity existence by ensuring a $\textit{Registered Event} (\text{WA}_c^c, \text{WA}_c^s, *)$ exists, where $\text{WA}_c^c$ is derived from $\text{SK}_c^c$, $\text{WA}_c^s$ from $\text{SK}_c^s$. If the validation result is $\bot$, terminate the protocol with \textit{Non-existence} exception. Otherwise, continue the protocol;
    \item Validate ownership by $\text{WA}_c^c = \textit{OwnerOf}(\iota)$, where $\textit{OwnerOf}$ is a method in the NFT protocol, to produce the validation result $\eta_0$;
    \item Validate request existence to produce $\eta_1$ by ensuring a $\textit{Requested Event} (\text{WA}_c^c, \iota)$ exists;
    \item Match the challenge message $M$ with the result of $\textit{Decrypt}(\hat{\text{PK}}, \rho, \chi_c)$ algorithm of the CP-ABE to produce $\eta_2$;
    \item Emit a $\textit{Verified Event}(\text{WA}_c^c, \iota, \eta)$ with $\text{SK}_c^c$, where $\eta = \eta_0 \lor \eta_1 \land \eta_2$.
\end{enumerate}

\paragraph{Decryption.}
This sub-protocol follows the \textit{Verification} protocol to decrypt $\textit{Enc}(data)$, which is also enforced by $\textit{Dec}$ in memory.

\begin{enumerate}
    \item Validate the verification result by ensuring a $\textit{Verified Event} (\text{WA}_c^c, \iota, \top)$ exists, where $\text{WA}_c^c$ is derived from $\text{SK}_c^c$;
    \item Get $\mu$ from $\textit{TokenURI}(\iota)$, a method in the NFT protocol;
    \item Get $\textit{Enc(D)}$ from $\mu[data] \implies S_{data}$ and verify its integrity by $\mathcal{H}(\textit{Enc(D)})$;
    \item Restore $\textit{Enc}(D)$ to $D$ with $\kappa$.
\end{enumerate}

\subsection{Multi-Chain Support}
\label{sec:multi_chain}
Sunspot allows a heterogeneous structure for a control chain and a storage chain. In a heterogeneous control chain, there is a master blockchain and a set of peer blockchains. And a storage chain can consist of a set of heterogeneous peer blockchains. In Sunspot, we formulate a Chain Manager and additional protocols to enable multi-chain compatibility on top of fundamental protocols illustrated in Section~\ref{sec:protocol}.

\subsubsection{Chain Manager}
Chain Manager is a DApp for registering information about blockchain clusters used to constitute the control chain or the storage chain. Its main functionality is to permanently store addresses of smart contracts deployed on all control-oriented blockchains and addresses of other components deployed on storage-oriented blockchains.

\subsubsection{Environment Assumption}
We assume the same environment illustrated in Section~\ref{sec:protocol_env} with additional assumptions as follows.
\begin{itemize}
    \item The control chain $\mathfrak{C}$ consists of a master blockchain $\mathfrak{M}$ and a finite set of peer blockchains $\textbf{N}_c = \{ \mathfrak{N}_0^c, \mathfrak{N}_1^c, \dots \}$.
    \item Chain Manager $C_{cm}$ is deployed on $\mathfrak{M}$ with a key pair $(\text{MPK}^m, \text{MSK}^m, \text{MWA}^m)$.
    \item The storage chain $\mathfrak{S}$ consists of a finite set of peer blockchains $\textbf{N}_s = \{ \mathfrak{N}_0^s, \mathfrak{N}_1^s, \dots \}$.
\end{itemize}

\subsubsection{Chain Initialization}
This protocol is executed on $\mathfrak{M}$ via $C_{cm}$ to register $\textbf{N}_c \cup \textbf{N}_s$. It contains two sub-protocols: \textit{Control Chain Registration} and \textit{Storage Chain Registration}.

\paragraph{Control Chain Registration.}
This sub-protocol registers all blockchains in $\textbf{N}_c$ to produce an address set $\textbf{C}$ of deployed contracts.

We present the main process for registering $\mathfrak{N}_i^c \in \textbf{N}_c$. This process is repeated for all elements in $\textbf{N}_c$.

\begin{enumerate}
    \item Get addresses $C_{dis}^i$ and $C_{id}^i$ by deploying Distributor and ID Manager optimized for $\mathfrak{N}_i$ to $\mathfrak{N}_i$ with $\text{MSK}_i^c$;
    \item Emit a $\textit{Control Registered Event} (i, \text{MWA}_i^c, C_{dis}^i, C_{id}^i)$ on $\mathfrak{M}$ via $C_{cm}$ with $\text{MSK}^m$;
    \item Append $(C_{dis}^i, C_{id}^i)$ to $\textbf{C}$.
\end{enumerate}

\paragraph{Storage Chain Registration.}
This sub-protocol registers all blockchains in $\textbf{N}_s$.

We present the main process for registering $\mathfrak{N}_j^s \in \textbf{N}_s$. This process is repeated for all elements in $\textbf{N}_s$.

\begin{enumerate}
    \item Configure and compile Cipher Suite, Control Linker, and Storage Linker with $\textbf{C}$. Store them to $\mathfrak{N}_j^s$ with $\text{MSK}_j^s$ and get addresses $S_{ciph}^j$, $S_{cl}^j$, $S_{sl}^j$ as their identifiers;
    \item Emit a $\textit{Storage Registered Event} (j, \text{MWA}_j^s, S_{ciph}^j, S_{cl}^j, S_{sl}^j)$ through $C_{cm}$ with $\text{MSK}^m$.
\end{enumerate}

\subsubsection{Chain Interaction}
For any protocol that has interactions with the control chain, it interacts with the target underlying blockchain through the Chain Manager $C_{cm}$.

Suppose a protocol calls a function of $C_{dis}^i$. After constructing $\textbf{C}$ by querying $\textit{Control Registered Events}$ of $\mathfrak{M}$, the protocol can identify the blockchain deploying $C_{dis}^i$ and invoke that function.

For interacting with the storage chain, it is the same with the single-chain case with the only difference of Application Programming Interfaces (APIs).

\section{MyPub}
\label{sec:mypub}
MyPub\footnote{https://github.com/yepengding/MyPub} is a decentralized privacy-preserving publishing platform based on Sunspot. Creators can publish their work such as digital paintings, writings, and music, without disclosing all contents on MyPub and share the metadata freely. To acquire the full contents of a work, customers can choose to buy the ownership or pay for the right to use. All operations of MyPub are enforced on blockchains via variants of protocols of Sunspot without central authorities. Based on Sunspot, MyPub can protect legal rights including the copyright, ownership, and right to use based on properties such as confidentiality, integrity, availability, and transparency.

MyPub adopts Ethereum as its control chain and Filecoin as its storage chain. To adapt to its business logic, MyPub simplifies but concretizes Sunspot protocols and adopts a management-free identity model that depends on a payment-based access control mechanism. This model does not require organizers to endorse participants' identities. Instead, it is based on the intrinsic transaction mechanism of the blockchain. For instance, to obtain the right to use of a work, the customer executes the $\textit{Requisition}$ protocol with a specific amount of cryptocurrency made by the owner. During the execution of the $\textit{Verification}$ protocol, the decryptor validates payment proofs, instead of authenticating the customer according to the registered attributes, to determine whether to decrypt the data. Payment proofs are forgery resistant because they have been already accepted in a blockchain network. Consequently, the payment-based access control mechanism is fully decentralized and does not require authorities for endorsements.

\section{Security and Privacy Analysis}
\subsection{Assumption}
We make reasonable assumptions below.

\paragraph{Blockchain Safety.}
Given a blockchain in Sunspot, most of the nodes, the number of which is greater than the threshold to make the consensus mechanism function well, are trustworthy. By assuming blockchain safety, any blockchain used in a system based on Sunspot preserves properties of a transparent public blockchain including on-chain data integrity and availability.

\paragraph{RPC Safety.}
Given a blockchain in Sunspot, its API at least enables TLS for RPCs. By assuming RPC safety, any blockchain used in a system based on Sunspot preserves connection integrity and RPC availability.

\paragraph{Control-Flow Integrity.}
Cipher Suite preserves control-flow integrity \cite{abadi_control-flow_2009}. Cipher Suite in Sunspot is implemented by the Rust \cite{matsakis_rust_2014}, a programming language that provides guarantees for memory safety \cite{jung_rustbelt_2017} through an ownership-based resource management model, without using \textit{unsafe} and \textit{external} blocks. Besides, the implementation follows the security through obscurity to protect sensitive function flows and constants.

\paragraph{Organizer Trust.}
By adopting $(g, |\textbf{O}|)$-threshold SSS, at least $g$ organizers out of $\textbf{O}$ are always available and constitute an available set $\textbf{O}_g \subseteq \textbf{O}$. $\forall o \in \textbf{O}_g$, $o$ can collude with at most $g'$ organizers where $g' < g - 1$. This assumption ensures there does not exist an organizer can dominate the identity management of a system based on Sunspot.

\subsection{Evaluation Model}
\label{sec:eval}
We use the same notation system illustrated in the sections above. To facilitate reading, we repeat some notations as follows.

For sets, we denote the participant set as $\textbf{P}$, the shared data set as $\textbf{D}$. Two types of blockchains in Sunspot are the control chain $\mathfrak{C}$ and the storage chain $\mathfrak{S}$.

\begin{property}[Data Privacy]
$\forall d \in \textbf{D}$, $p \in \textbf{P}$ can decrypt $d$ if and only if $\eta^p = \top$.
\begin{proof}
In Sunspot, data privacy is ensured by confidentiality. Data are encrypted through the \textit{Encryption} protocol before being stored on $\mathfrak{S}$, which introduces confidentiality for the stored data.

\paragraph{$\Rightarrow$}
In the \textit{Encryption} protocol, $d$ is encrypted via the AES with secret key $\kappa$ obfuscated and hard-coded in its decryptor. By the \textit{Control-Flow Integrity} assumption, $\kappa$ is not disclosed to any $p \in \textbf{P}$, of which the safety is guaranteed. Hence, executing the \textit{Authorization} protocol is the only way for $p$ to decrypt $d$. In the sub-protocol \textit{Verification}, $\eta^p$ is the only condition to determine the enforcement of the \textit{Decryption} protocol. Besides, $\eta^p$ becomes tamper-evident and immutable after corresponding \textit{Verified Event} is emitted according to the \textit{Blockchain Safety}. If $p$ can decrypt $d$, $\eta^p$ must be satisfied.

\paragraph{$\Leftarrow$}
In the \textit{Verification} protocol, $\eta^p = \top$ holds in two cases: $\eta_0^p = \top$ or $\bigwedge\limits_{i=1}^2 \eta_i^p = \top$. The first case implies $p$ is the owner of $d$, while the second case implies the attributes of $p$ satisfies the access control policy $\rho_d$, which is ensured by the CP-ABE. Based on the assumptions above, $p$ can decrypt $d$ in either case when $\eta^p = \top$.
\end{proof}
This completes the proof.
\end{property}

\begin{property}[Identity Privacy]
$\forall p \in \textbf{P}$ with a set of attributes $\textbf{A}_p$, $\textbf{A}_p$ cannot be disclosed.

\begin{proof}
Sunspot protects the privacy of identities in all types of identity management models. Since identity management depends on its underlying access control mechanisms, two cases in Sunspot are proved below.

\begin{case}[Fine-Grained Mechanism]
In the \textit{Registration} protocol, the private key $\chi_p$ that describes $\textbf{A}_p$ is generated based on the CP-ABE and persist on $\mathfrak{C}$. $\chi_p$ is the only public information associated with $\textbf{A}_p$. Since $\textbf{A}_p$ cannot be disclosed with  $\chi_p$ according to \cite{waters_ciphertext-policy_2011}, the identity privacy is protected.
\end{case}

\begin{case}[Payment-Based Mechanism]
This mechanism does not require identity information other than the wallet address embedded in a payment transaction. Since the wallet address does not directly link to the real identity, the identity information cannot be disclosed.
\end{case}
This completes the proof.
\end{proof}
\end{property}

\begin{property}[Data Availability]
$\forall d \in \textbf{D}$, if the token id $\iota$ of $d$ is known, $d$ is always available.

\begin{proof}
The metadata $\mu$ of $d$ is always available via calling $\textit{TokenURI}(\iota)$ of the Distributor $C_{dis}$ according to the \textit{Blockchain Safety} and \textit{RPC Safety} assumptions. On-chain data integrity implies $\mu$ and $S_{data}$ implied from $\mu[data]$ are tamper-proof. Based on the \textit{Blockchain Safety}, $d$ is always available via its address $S_{data}$.
\end{proof}
\end{property}

\subsection{Experiments}
To demonstrate the properties are effective to protect Sunspot from threats, we conduct two representative experiments from the perspective of attackers in a simulation environment where we set up an Ethereum and a Filecoin network with TLS enabled, as well as 10 account pairs (one for each network) for organizers, 10 for data providers, and 10 for data consumers. The \textit{Initialization} and \textit{Registration} protocols are executed after the environment setup.

\subsubsection{Man-in-the-Middle Attack}
We simulate a middleman who aims to deceive the verification process of the decryptor by hijacking the connections between the decryptor and the blockchain network. After penetrating our deliberately developed vulnerable network, all packets are successfully intercepted by the middle man. However, without the private key, the middle man can neither know the transmitting data nor forge a connection to compromise the $\textit{Verification}$ protocol.

We also formulate an advanced attack by making the middleman pretend to be a legal node in the network with a self-signed certificate. However, the decryptor throws an \textit{Unauthenticated Error} immediately after detecting an unknown certificate authority in the first process of the \textit{Verification} protocol.

\subsubsection{Return-Oriented Programming}
To test the control-flow integrity, we use return-oriented programming (ROP) \cite{roemer_return-oriented_2012} to analyze the decryptor and try to find gadgets to bypass the \textit{Verification} protocol by modifying the control flow. We decode the decryptor into a file in the LLVM assembly language format \cite{lattner_llvm_2004}. However, the file is over 2 million lines with unreadable flows and values due to the obfuscation. And we fail to exploit it with static and runtime analyzers such as \cite{kroustek_retdec_2017}.

\section{Performance Analysis}
In Sunspot, we evaluate the performance of protocols regarding a set of data sizes. For each data size per protocol, we conducted experiments 10 times. To evaluate the performance without invasion, we use containerization techniques to encapsulate components to minimize the impact of irrelevant factors such as network delay. We also implement external scripts to observe the performance in a cloud environment (Amazon EC2 g4dn.xlarge). The result is shown in Table~\ref{tab:perf}.

Particularly, we assume a constant time for on-chain operations that are irrelevant factors (e.g., mining and querying) to the performance evaluation of Sunspot, though on-chain operations probably cause a significant performance impact in practice. We implement this by subtracting the observed on-chain operation time from raw results.

\begin{table}
\caption{The performance of Sunspot protocols.}\label{tab:perf}
\begin{tabular}{|l|c|c|c|c|}
\hline
\textbf{Data size}&\multicolumn{4}{|c|}{\textbf{Protocols (s)}} \\
\cline{2-5} 
\textbf{(byte)} & \textbf{\textit{Initialization}}& \textbf{\textit{Registration}}& \textbf{\textit{Solidification}} & \textbf{\textit{Authorization}} \\
\hline
1,024 & 17.53 & 0.36 & 9.47 & 3.26 \\
2,048 &  13.62 & 0.35 & 9.99 & 4.01 \\
4,096 & 11.15 & 0.38 & 11.03 & 4.87 \\
8,192 & 13.56 & 0.44 & 13.60 & 5.89 \\
16,384 & 14.01 & 0.35 & 17.85 & 7.68 \\
\hline
\end{tabular}
\end{table}

Notably, the \textit{Initialization} protocol includes the compiling time of components, which brings a considerable time cost. The same with the \textit{Encryption} sub-protocol of the \textit{Solidification} that generates and compiles a decryptor in the execution process.

\section{Discussion}
\label{sec:discussion}
We claim that Sunspot provides a feasible solution to the issues and challenges described in Section~\ref{sec:intro}.

\paragraph{Challenge 1}[Proof of legal rights]
Metadata are stored on the control chain as NFTs. The proofs of legal rights derived from on-chain data are immutably verifiable due to the public blockchain properties. For instance, in Mypub, a creator of an NFT holds copyright and initial ownership while a customer can hold the right to use of some data after the payment, i.e., getting authorized. These rights are permanently traceable and verifiable.

\paragraph{Challenge 2}[Fairly large data storage]
In Sunspot, the control chain collaborates with the storage chain to enable fairly large data storage. The storage chain is implemented by decentralized storage systems to store large data. Besides, a cross-chain collaboration mechanism is implemented based on protocols in Section~\ref{sec:protocol} to bridge the control chain and the storage chain.

\paragraph{Challenge 3}[Interoperability]
First, Sunspot protocols and access control mechanisms are generalized and not blockchain-specific. Therefore, Sunspot can support all types of blockchains that satisfy the conditions defined in Section~\ref{sec:sunspot}. Furthermore, Sunspot enables multi-chain compatibility and allows heterogeneous structures by the Chain Manager and additional protocols illustrated in Section~\ref{sec:multi_chain}. In this manner, a chain component can consist of multiple heterogeneous blockchains to be scalable and dependable.

\paragraph{Challenge 4}[Privacy preservation]
In Sunspot, identity privacy and data privacy are preserved. We provide proofs in Section~\ref{sec:eval}. Participants' attributes are stored on chain but described by a private key generated in the \textit{Registration} protocol. And the shared data are fully encrypted via the \textit{Encryption} protocol before being stored on the storage chain.

\paragraph{Challenge 5}[Access control]
Currently, two access control mechanisms are mechanized in Sunspot including a fine-grained mechanism and a payment-based mechanism. Based on these two mechanisms, Sunspot supports an identity management model based on the combination of the CP-ABE and blockchains that is described in Section~\ref{sec:sunspot} and a management-free model purely enforced on the blockchain that is described in Section~\ref{sec:mypub}.

However, the identity management model based on the fine-grained access control mechanism highly depends on the \textit{Control-Flow Integrity} assumption. Although the management-free model is enforced on the blockchain, it lacks support for complex access control policies. Therefore, the implementation of the self-sovereign identity (SSI) \cite{muhle_survey_2018} model is on our schedule. With the support of the SSI, participants can manage their own identities based on blockchains and present on-demand verifiable presentations for authentication and authorization.

Besides, Sunspot-based systems are also threatened by the security issues of blockchains such as smart contract vulnerabilities \cite{destefanis_smart_2018} and the Sybil attack \cite{zhang_double-spending_2019}.

\section{Conclusion}
In this paper, we have presented Sunspot, a decentralized privacy-preserving framework with multiple access control mechanisms, to solve the key challenges of data sharing on transparent public blockchains including proof of legal rights, fairly large data storage, interoperability, privacy preservation, and access control. To enlarge the applicability of Sunspot, we have generalized its protocols and made it support multi-chain and heterogeneous structures. Besides, we have proved the security and privacy properties of Sunspot including data privacy, identity privacy, and data availability. For security in practice, we have conducted experiments from the perspective of attackers to launch representative attacks to demonstrate the effectiveness of Sunspot. We have also evaluated the performance of core functionalities implemented by formulated protocols. Furthermore, we have discussed the methods mechanized in Sunspot of solving the key challenges of data sharing on transparent public blockchains, as well as the limitations and further improvements.

%
%
%
\bibliographystyle{splncs04}
\bibliography{sunspot}

\begin{thebibliography}{10}
\providecommand{\url}[1]{\texttt{#1}}
\providecommand{\urlprefix}{URL }
\providecommand{\doi}[1]{https://doi.org/#1}

\bibitem{abadi_control-flow_2009}
Abadi, M., Budiu, M., Erlingsson, U., Ligatti, J.: Control-flow integrity
  principles, implementations, and applications. ACM Transactions on
  Information and System Security (TISSEC)  \textbf{13}(1),  1--40 (2009),
  iSBN: 1094-9224 Publisher: ACM New York, NY, USA

\bibitem{androulaki_hyperledger_2018}
Androulaki, E., Barger, A., Bortnikov, V., Cachin, C., Christidis, K., De~Caro,
  A., Enyeart, D., Ferris, C., Laventman, G., Manevich, Y.: Hyperledger fabric:
  a distributed operating system for permissioned blockchains. In: Proceedings
  of the {Thirteenth} {EuroSys} {Conference}. pp. 1--15 (2018)

\bibitem{bartoletti_sok_2021}
Bartoletti, M., Chiang, J.H.y., Lafuente, A.L.: {SoK}: lending pools in
  decentralized finance. In: International {Conference} on {Financial}
  {Cryptography} and {Data} {Security}. pp. 553--578. Springer (2021)

\bibitem{benet_ipfs-content_2014}
Benet, J.: Ipfs-content addressed, versioned, p2p file system. arXiv preprint
  arXiv:1407.3561  (2014)

\bibitem{chen_blockchain_2020}
Chen, Y., Bellavitis, C.: Blockchain disruption and decentralized finance:
  {The} rise of decentralized business models. Journal of Business Venturing
  Insights  \textbf{13},  e00151 (2020), iSBN: 2352-6734 Publisher: Elsevier

\bibitem{dang_secure_2015}
Dang, Q.H.: Secure hash standard  (2015)

\bibitem{destefanis_smart_2018}
Destefanis, G., Marchesi, M., Ortu, M., Tonelli, R., Bracciali, A., Hierons,
  R.: Smart contracts vulnerabilities: a call for blockchain software
  engineering? In: 2018 {International} {Workshop} on {Blockchain} {Oriented}
  {Software} {Engineering} ({IWBOSE}). pp. 19--25. IEEE (2018)

\bibitem{ding_derepo_2020}
Ding, Y., Sato, H.: Derepo: a distributed privacy-preserving data repository
  with decentralized access control for smart health. In: 2020 7th {IEEE}
  {International} {Conference} on {Cyber} {Security} and {Cloud} {Computing}
  ({CSCloud})/2020 6th {IEEE} {International} {Conference} on {Edge}
  {Computing} and {Scalable} {Cloud} ({EdgeCom}). pp. 29--35. IEEE (2020)

\bibitem{dworkin_sha-3_2015}
Dworkin, M.J.: {SHA}-3 standard: {Permutation}-based hash and extendable-output
  functions  (2015)

\bibitem{entriken_erc-721_2018}
Entriken, W., Shirley, D., Evans, J., Sachs, N.: Erc-721 non-fungible token
  standard. Ethereum Foundation  (2018)

\bibitem{garay_bitcoin_2015}
Garay, J., Kiayias, A., Leonardos, N.: The bitcoin backbone protocol:
  {Analysis} and applications. In: Annual international conference on the
  theory and applications of cryptographic techniques. pp. 281--310. Springer
  (2015)

\bibitem{goldreich_definitions_1994}
Goldreich, O., Oren, Y.: Definitions and properties of zero-knowledge proof
  systems. Journal of Cryptology  \textbf{7}(1),  1--32 (1994), iSBN: 1432-1378
  Publisher: Springer

\bibitem{goldwasser_best-possible_2007}
Goldwasser, S., Rothblum, G.N.: On best-possible obfuscation. In: Theory of
  {Cryptography} {Conference}. pp. 194--213. Springer (2007)

\bibitem{groth_updatable_2018}
Groth, J., Kohlweiss, M., Maller, M., Meiklejohn, S., Miers, I.: Updatable and
  universal common reference strings with applications to zk-{SNARKs}. In:
  Annual {International} {Cryptology} {Conference}. pp. 698--728. Springer
  (2018)

\bibitem{gueron_aes-gcm-siv_2017}
Gueron, S., Langley, A., Lindell, Y.: {AES}-{GCM}-{SIV}: {Specification} and
  {Analysis}. IACR Cryptol. ePrint Arch.  \textbf{2017}, ~168 (2017)

\bibitem{johnson_elliptic_2001}
Johnson, D., Menezes, A., Vanstone, S.: The elliptic curve digital signature
  algorithm ({ECDSA}). International journal of information security
  \textbf{1}(1),  36--63 (2001), publisher: Springer

\bibitem{jung_rustbelt_2017}
Jung, R., Jourdan, J.H., Krebbers, R., Dreyer, D.: {RustBelt}: {Securing} the
  foundations of the {Rust} programming language. Proceedings of the ACM on
  Programming Languages  \textbf{2}(POPL),  1--34 (2017), iSBN: 2475-1421
  Publisher: ACM New York, NY, USA

\bibitem{kroustek_retdec_2017}
Křoustek, J., Matula, P., Zemek, P.: Retdec: {An} open-source machine-code
  decompiler. December (2017)

\bibitem{lattner_llvm_2004}
Lattner, C., Adve, V.: {LLVM}: {A} compilation framework for lifelong program
  analysis \& transformation. In: International {Symposium} on {Code}
  {Generation} and {Optimization}, 2004. {CGO} 2004. pp. 75--86. IEEE (2004)

\bibitem{liu_linkable_2004}
Liu, J.K., Wei, V.K., Wong, D.S.: Linkable spontaneous anonymous group
  signature for ad hoc groups. In: Australasian {Conference} on {Information}
  {Security} and {Privacy}. pp. 325--335. Springer (2004)

\bibitem{matsakis_rust_2014}
Matsakis, N.D., Klock, F.S.: The rust language. ACM SIGAda Ada Letters
  \textbf{34}(3),  103--104 (2014), iSBN: 1094-3641 Publisher: ACM New York,
  NY, USA

\bibitem{miers_zerocoin_2013}
Miers, I., Garman, C., Green, M., Rubin, A.D.: Zerocoin: {Anonymous}
  distributed e-cash from bitcoin. In: 2013 {IEEE} {Symposium} on {Security}
  and {Privacy}. pp. 397--411. IEEE (2013)

\bibitem{muhle_survey_2018}
Mühle, A., Grüner, A., Gayvoronskaya, T., Meinel, C.: A survey on essential
  components of a self-sovereign identity. Computer Science Review
  \textbf{30},  80--86 (2018), iSBN: 1574-0137 Publisher: Elsevier

\bibitem{nakamoto_bitcoin_2019}
Nakamoto, S.: Bitcoin: {A} peer-to-peer electronic cash system. Tech. rep.,
  Manubot (2019)

\bibitem{noether_ring_2015}
Noether, S.: Ring {SIgnature} {Confidential} {Transactions} for {Monero}. IACR
  Cryptol. ePrint Arch.  \textbf{2015}, ~1098 (2015)

\bibitem{provos_bcrypt_1999}
Provos, N., Mazieres, D.: Bcrypt algorithm. In: {USENIX} (1999)

\bibitem{regulation_regulation_2016}
Regulation, G.D.P.: Regulation {EU} 2016/679 of the {European} {Parliament} and
  of the {Council} of 27 {April} 2016. Official Journal of the European Union
  (2016)

\bibitem{roemer_return-oriented_2012}
Roemer, R., Buchanan, E., Shacham, H., Savage, S.: Return-oriented programming:
  {Systems}, languages, and applications. ACM Transactions on Information and
  System Security (TISSEC)  \textbf{15}(1),  1--34 (2012), iSBN: 1094-9224
  Publisher: ACM New York, NY, USA

\bibitem{schwartz_ripple_2014}
Schwartz, D., Youngs, N., Britto, A.: The ripple protocol consensus algorithm.
  Ripple Labs Inc White Paper  \textbf{5}(8), ~151 (2014)

\bibitem{shamir_how_1979}
Shamir, A.: How to share a secret. Communications of the ACM  \textbf{22}(11),
  612--613 (1979), iSBN: 0001-0782 Publisher: ACm New York, NY, USA

\bibitem{sun_ringct_2017}
Sun, S.F., Au, M.H., Liu, J.K., Yuen, T.H.: Ringct 2.0: {A} compact
  accumulator-based (linkable ring signature) protocol for blockchain
  cryptocurrency monero. In: European {Symposium} on {Research} in {Computer}
  {Security}. pp. 456--474. Springer (2017)

\bibitem{van_saberhagen_cryptonote_2013}
Van~Saberhagen, N.: {CryptoNote} v 2.0 (2013)

\bibitem{vazirani_efficient_1984}
Vazirani, U.V., Vazirani, V.V.: Efficient and secure pseudo-random number
  generation. In: Workshop on the {Theory} and {Application} of {Cryptographic}
  {Techniques}. pp. 193--202. Springer (1984)

\bibitem{wan_pride_2018}
Wan, Z., Guan, Z., Cheng, X.: Pride: {A} private and decentralized usage-based
  insurance using blockchain. In: 2018 {IEEE} {International} {Conference} on
  {Internet} of {Things} ({iThings}) and {IEEE} {Green} {Computing} and
  {Communications} ({GreenCom}) and {IEEE} {Cyber}, {Physical} and {Social}
  {Computing} ({CPSCom}) and {IEEE} {Smart} {Data} ({SmartData}). pp.
  1349--1354. IEEE (2018)

\bibitem{waters_ciphertext-policy_2011}
Waters, B.: Ciphertext-policy attribute-based encryption: {An} expressive,
  efficient, and provably secure realization. In: International {Workshop} on
  {Public} {Key} {Cryptography}. pp. 53--70. Springer (2011)

\bibitem{wood_ethereum_2014}
Wood, G.: Ethereum: {A} secure decentralised generalised transaction ledger.
  Ethereum project yellow paper  \textbf{151}(2014),  1--32 (2014)

\bibitem{wu_cloud_2010}
Wu, J., Ping, L., Ge, X., Wang, Y., Fu, J.: Cloud storage as the infrastructure
  of cloud computing. In: 2010 {International} {Conference} on {Intelligent}
  {Computing} and {Cognitive} {Informatics}. pp. 380--383. IEEE (2010)

\bibitem{xia_bbds_2017}
Xia, Q., Sifah, E.B., Smahi, A., Amofa, S., Zhang, X.: {BBDS}:
  {Blockchain}-based data sharing for electronic medical records in cloud
  environments. Information  \textbf{8}(2), ~44 (2017), publisher:
  Multidisciplinary Digital Publishing Institute

\bibitem{zetzsche_decentralized_2020}
Zetzsche, D.A., Arner, D.W., Buckley, R.P.: Decentralized finance. Journal of
  Financial Regulation  \textbf{6}(2),  172--203 (2020), iSBN: 2053-4841
  Publisher: Oxford University Press

\bibitem{zhang_blockchain-based_2018}
Zhang, G., Li, T., Li, Y., Hui, P., Jin, D.: Blockchain-based data sharing
  system for ai-powered network operations. Journal of Communications and
  Information Networks  \textbf{3}(3), ~1--8 (2018), iSBN: 2509-3312 Publisher:
  Springer

\bibitem{zhang_double-spending_2019}
Zhang, S., Lee, J.H.: Double-spending with a sybil attack in the bitcoin
  decentralized network. IEEE transactions on Industrial Informatics
  \textbf{15}(10),  5715--5722 (2019), iSBN: 1551-3203 Publisher: IEEE

\end{thebibliography}
\end{document}